\documentclass[12pt,reqno]{amsart}

\usepackage{amssymb,latexsym,graphicx}

\newtheorem{theorem}{Theorem}[section]
\newtheorem{lemma}[theorem]{Lemma}
\newtheorem{proposition}[theorem]{Proposition}

\theoremstyle{definition}

\theoremstyle{remark}
\newtheorem{remark}[theorem]{Remark}

\numberwithin{equation}{section}

    \renewcommand{\Re}{{\operatorname{Re}}}
    \renewcommand{\Im}{{\operatorname{Im}}}
    \DeclareMathOperator{\Ai}{Ai}
    \DeclareMathOperator{\bigO}{{\mathcal O}}

    \newcommand{\supN}{(\hspace{-0.04cm}N\hspace{-0.04cm})}

    \DeclareMathOperator{\supp}{supp}
    \DeclareMathOperator{\sgn}{sgn}
    
    \DeclareMathOperator*{\Tr}{Tr}

\begin{document}
\title[When the soft edge meets the hard edge]{Universality in unitary
random matrix ensembles  when the soft edge meets the hard edge}
\author{Tom Claeys and Arno B.J. Kuijlaars}
\address{Department of Mathematics,
Katholieke Universiteit Leuven,
Celestijnenlaan 200B,
3001 Leuven, Belgium}
\email{tom.claeys@wis.kuleuven.be, arno.kuijlaars@wis.kuleuven.be}


\subjclass[2000]{Primary 15A52, 34M55; Secondary 35Q15}

\dedicatory{Dedicated to Percy Deift on the occasion of his
sixtieth birthday}
\thanks{The authors are supported by FWO-Flanders project G.0455.04,
by K.U. Leuven research grant OT/04/21,
by the Belgian Interuniversity Attraction Pole P06/02, and by the
 European Science Foundation Program MISGAM. The second author is
 also supported by INTAS Research Network 03-51-6637,
and by a grant from the Ministry of Education and
Science of Spain, project code MTM2005-08648-C02-01.}

\begin{abstract}
Unitary random matrix ensembles \[Z_{n,N}^{-1} (\det M)^{\alpha}
\exp(-N \Tr V(M))\, dM\] defined on positive definite matrices
$M$, where $\alpha > -1$ and $V$ is real analytic, have a hard
edge at $0$. The equilibrium measure associated with $V$ typically
vanishes like a square root at soft edges of the spectrum. For the
case that the equilibrium measure vanishes like a square root at
$0$, we determine the scaling limits of the eigenvalue correlation
kernel near $0$ in the limit when $n, N \to \infty$ such that $n/N
- 1 = \bigO(n^{-2/3})$. For each value of $\alpha > -1$ we find a
one-parameter family of limiting kernels that we describe in terms
of the Hastings-McLeod solution of the Painlev\'e II equation with
parameter $\alpha + 1/2$.
\end{abstract}

\maketitle

\section{Introduction and statement of results}

We deal in this paper with the following question.

\begin{description}
\item[Question]
Given a unitary random matrix ensemble
\begin{equation} \label{ensemble1} \frac{1}{Z_{n,N}}
(\det M)^{\alpha} \exp(-N \Tr V(M)) \, dM,
    \qquad \alpha > -1,
\end{equation}
defined on positive definite Hermitian matrices $M$ of size $n \times n$,
where the real analytic potential $V$ is such that the equilibrium measure
associated with $V$ has a density on $[0,\infty)$ that vanishes like
a square root at $0$.
What are the scaling limits around $0$ of the eigenvalue correlation kernel
as $n, N \to \infty$, $n/ N \to 1$ ?
\end{description}

One expects a universality type result for this situation where
the limiting kernels do not depend on the exact form of $V$.
This is in line with known universality results in the usual
unitary random matrix ensembles of the form
\begin{equation} \label{ensemble2}
    \frac{1}{Z_{n,N}} |\det M|^{2\alpha} \exp(-N \Tr W(M)) \, dM,
        \qquad \alpha > -1/2
\end{equation}
defined on all Hermitian matrices $M$ of size $n \times n$,
\cite{BI1,BI2,CK,CKV,DKMVZ1,KV}.
In (\ref{ensemble1}) and (\ref{ensemble2}) and throughout the
paper we assume that the confining potentials $V$ and $W$ are
real analytic and satisfy
\begin{equation} \label{asympVW}
    \lim_{x \to +\infty} \frac{V(x)}{\log(x^2+1)} = + \infty,
    \qquad \mbox{ and } \qquad
    \lim_{x \to \pm \infty} \frac{W(x)}{\log(x^2+1)} = + \infty. \end{equation}
The random matrix ensemble (\ref{ensemble1}) is restricted to
positive definite matrices, thereby creating a hard edge at $0$.
This has an influence on the local eigenvalue behavior near $0$
since eigenvalues are always positive.

It is well-known, see e.g.~\cite{Deift,Mehta}, that the eigenvalue
correlation kernel for the ensemble (\ref{ensemble1}) has the form
\begin{equation}\label{kernel1}
    K_{n,N}(x,y) = x^\frac{\alpha}{2} y^\frac{\alpha}{2} e^{-\frac{N}{2}V(x)}e^{-\frac{N}{2}V(y)}
    \sum_{j=0}^{n-1}p_j^{\supN}(x)p_j^{\supN}(y),
\end{equation}
where $p_j^{\supN}$ is the $j$-th degree orthonormal polynomial
with respect to the weight $x^\alpha e^{-NV(x)}$ on $[0,\infty)$.
The limiting mean eigenvalue density
\[ \psi_V(x) = \lim_{n,N \to \infty \atop n/N \to 1} \frac{1}{n} K_{n,N}(x,x) \]
is the minimizer of the weighted energy
\begin{equation} \label{energy1}
I_V(\psi)=\iint \log \frac{1}{|x-y|} \, \psi(x) \psi(y)\, dxdy
    + \int V(x) \, \psi(x) \, dx
\end{equation}
taken over probability density functions $\psi$ supported on $[0,\infty)$,
see \cite{Deift,ST}. It does not depend on $\alpha$. As $V$ is real analytic,
the support of $\psi_V$ consists of a finite union of intervals \cite{DKM}.

\begin{figure}[t]
\begin{center}
\includegraphics[angle=270]{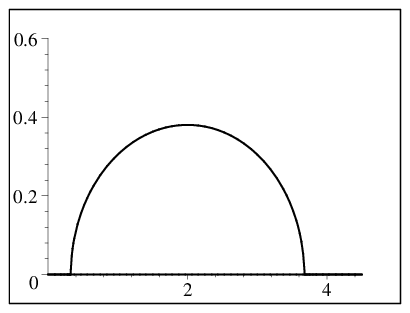} \
\includegraphics[angle=270]{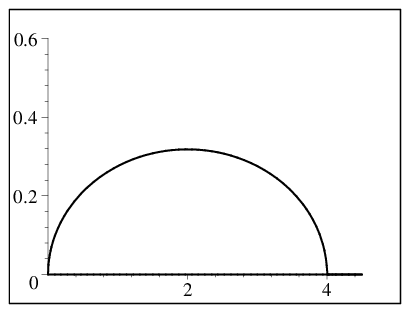} \
\includegraphics[angle=270]{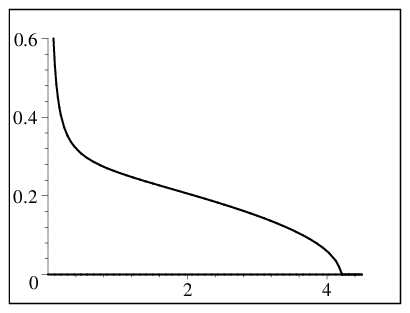}
\end{center}
\caption{The density $\psi_{V_c}$ for $V_c(x)= \frac{1}{2c}(x-2)^2$
with $c$ equal to $0.7$ (left figure), $1$ (middle figure) and $1.2$ (right figure).
\label{figure1}}
\end{figure}

We can then distinguish two situations. The first is that $0$
does not belong to the support $S_V$ of $\psi_V$. Then the hard edge at
$0$ does not lead to a new phenomenon. Generically one finds
that the eigenvalue correlation kernel (\ref{kernel1}) has the sine kernel
\begin{equation} \label{sinekernel}
    \frac{\sin \pi(x-y)}{\pi(x-y)}
    \end{equation}
as scaling limit in the bulk and the Airy kernel
\begin{equation} \label{Airykernel}
    \frac{\Ai(x) \Ai'(y) - \Ai'(x) \Ai(y)}{x-y}
\end{equation}
as scaling limit at the (soft) edge points of $S_V$.
These are also the generic scaling limits for the correlation kernel
associated with the matrix ensemble (\ref{ensemble2}), see \cite{BI1,Deift,DG,DKMVZ1,TW1}.

The second situation is that $0$ belongs to $S_V$.
In that case, one typically finds that $\psi_V$ is unbounded
at $0$ with a square root type singularity at $0$.
Then instead of the Airy kernel we are led to a Bessel kernel
\begin{equation} \label{Besselkernel}
    \frac{J_{\alpha}(\sqrt x)\sqrt y J_{\alpha}'(\sqrt y)-J_{\alpha}(\sqrt y)\sqrt x J_{\alpha}'(\sqrt x)}{2(x-y)}
\end{equation}
that describes the local eigenvalue correlations near $0$,
see \cite{KV,TW2,Vanl}.

Changing parameters in $V$ we may create a transition from the
Airy kernel to the Bessel kernel. This happens for example if
\[ V(x) = V_c(x) = \frac{1}{2c}(x-2)^2. \]
When considered on $\mathbb R$ this is a shifted
and scaled  GUE ensemble. For $c < 1$ we have
\[ \psi_{V_c}(x) = \frac{1}{2 \pi c} \sqrt{4c -(x-2)^2},
    \qquad \mbox{for } x \in [2 - 2\sqrt{c}, 2+2\sqrt{c}]. \]
These are all semi-circle laws which tend as $c \to 1^{-}$
to the semi-circle law on $[0,4]$. Then the left soft edge
meets with the hard edge at $0$. For $c > 1$ we have
\[ \psi_{V_c}(x) = \frac{1}{2\pi c \sqrt{x}}
    (x + a) \sqrt{b-x},
        \qquad \mbox{for } x \in [0,b], \]
with
\[ a = -\frac{4}{3} + \frac{2}{3} \sqrt{1+3c} \qquad
    \mbox{ and } \qquad b = \frac{4}{3} + \frac{4}{3}\sqrt{1+3c}, \]
which
has a square-root singularity at $0$. See Figure \ref{figure1}
for the density $\psi_{V_c}$ for the values $c=0.7$, $c=1$ and $c=1.2$.

Thus our Question asks about the transitional case. What happens
when the soft edge meets the hard edge? Note that this transition
from soft to hard edge is  different from the one considered in
\cite{BF} where the authors start from a hard edge situation and
let the parameter $\alpha$ increase to infinity. Other types
of singular edge behaviors in unitary random matrix ensembles have
been studied in \cite{IKO}, where a soft edge coincides with a spectral
singularity, and in \cite{CV}, where the limiting mean eigenvalue density
vanishes faster than in the regular case.

\begin{remark}
Our Question is related to the calculation of the
so-called Janossy densities for unitary random matrix ensembles
\[ \frac{1}{Z_{n,N}} \exp(-N \Tr V(M)) \, dM \]
in an asymptotic limit.
Indeed, it was shown in \cite{BS} that the Janossy densities
for an interval $I \subset \mathbb R$
are expressed in terms of  the orthogonal polynomials for the weight
$e^{-N V(x)}$ restricted to the complement of $I$.
If $\psi_V$ vanishes like a square root at a left edge point $a$,
and if $I = (-\infty,a]$, then after linear scaling we
can reduce this to the case $a =0$, and then the Janossy
densities are expressed in terms of the
orthogonal polynomial kernel (\ref{kernel1}) with $\alpha = 0$.
\end{remark}

Our first result is that there is indeed a one-parameter family of limiting kernels
that depends on $\alpha$.

\begin{theorem} \label{theorem11}
For every $\alpha > -1$, there is a one-parameter family of kernels
$\mathbb K_{\alpha}^{\rm soft/hard}(x,y;s)$ depending on $s \in \mathbb R$,
such that the following holds. Let $V$ be real analytic on $[0,\infty)$ such that $\psi_V$
vanishes like a square root at $0$ and such that there are no other singular points
in the external field $V$. Then there are positive constants $c = c_{1,V}$ and $c_{2,V}$ such
that
\begin{equation} \label{limitkernel}
    \lim_{n,N \to \infty} \frac{1}{(cn)^{2/3}} K_{n,N}\left( \frac{x}{(cn)^{2/3}}, \frac{y}{(cn)^{2/3}} \right)
    = \mathbb K_{\alpha}^{\rm soft/hard}(x,y; s) \end{equation}
whenever $n, N \to \infty$ such that
\begin{equation} \label{eq-limits}
    \lim\limits_{n\to \infty} n^{2/3} \left( \frac{n}{N} - 1\right) = L \in \mathbb R
\end{equation}
and $s = c_{2,V} L$.
The limit {\rm (\ref{limitkernel})} holds
uniformly for $x$ and $y$ in compact subsets of $(0,\infty)$.
\end{theorem}
See equation (\ref{defc1Vc2V}) below for explicit formulas for the
constants $c_{1,V}$ and $c_{2,V}$. We also refer to \cite{DKMVZ1}
and  Section 3 below for the notion and the classification of
singular points in an external field.

In contrast to other works on universality, see e.g.\ \cite{CKV,DKMVZ1},
we do not have to go through an explicit Riemann-Hilbert steepest descent
analysis in order to prove Theorem \ref{theorem11}. Instead we prove
Theorem \ref{theorem11} by relating the random matrix ensemble (\ref{ensemble1})
to the random matrix ensemble (\ref{ensemble2}) defined on all
$n \times n$ Hermitian matrices  where
\[ W(x) = \frac{1}{2} V(x^2) \qquad \mbox{ for } x \in \mathbb R \]
and with $\alpha$ replaced by $\alpha \pm 1/2$. The equilibrium measure
$\psi_W$ for (\ref{ensemble2}) vanishes quadratically at $0$. The
universal limiting kernels for this situation were found in \cite{CKV}.
They are constructed out of $\psi$-functions associated with
the Hastings-McLeod solution of the Painlev\'e II equation. Using
this connection we find the following representation for
the limiting kernels $\mathbb K_{\alpha}^{\rm soft/hard}(x,y;s)$.

\begin{theorem} \label{theorem12}
Let $\alpha > -1$. Then the kernels $\mathbb K_{\alpha}^{\rm soft/hard}(x,y;s)$
have the form
\begin{equation} \label{integrablekernel}
    \mathbb K_{\alpha}^{\rm soft/hard}(x,y; s) =
    \frac{f_{\alpha}(x;s) g_{\alpha}(y;s) - f_{\alpha}(y;s) g_{\alpha}(x;s)}{\pi(x-y)}
    \end{equation}
where $f_{\alpha}$ and $g_{\alpha}$ are  solutions of the system of differential equations
\begin{equation} \label{diffsystem}
    \frac{d}{dx} \begin{pmatrix} f_{\alpha}(x;s) \\[5pt] g_{\alpha}(x;s) \end{pmatrix}
    = \begin{pmatrix} 2q + \alpha/(2x) & 2 + (q^2 + r + \frac{s}{2})/x \\[5pt]
    -2 x - q^2 + r - \frac{s}{2} & -2q - \alpha/(2x) \end{pmatrix}
    \begin{pmatrix} f_{\alpha}(x;s) \\[5pt] g_{\alpha}(x;s) \end{pmatrix}
    \end{equation}
where $q=q(s)$ is the Hastings-McLeod solution of the Painlev\'e II equation
with parameter $\alpha + \frac{1}{2}$
\begin{equation} \label{Painleve2}
    q'' = sq + 2 q^3 - \alpha - \frac{1}{2}
    \end{equation}
and $r = r(s) = q'(s)$.
The solutions of {\rm (\ref{diffsystem})} are characterized by the asymptotic behavior
\begin{align} \label{realfalphaasymptotics}
    f_{\alpha}(x;s) & =  x^{-1/4}
        \cos\left(\frac{4}{3}x^{3/2}+sx^{1/2} - \frac{1}{2}\pi (\alpha+\frac{1}{2})\right)
     + \bigO(x^{-3/4}), \\
    \label{realgalphaasymptotics}
    g_{\alpha}(x;s) & = - x^{1/4}
        \sin\left(\frac{4}{3}x^{3/2}+sx^{1/2} - \frac{1}{2}\pi (\alpha+\frac{1}{2})\right)
    + \bigO(x^{-1/4}), \end{align}
as $x \to +\infty$.
\end{theorem}

\begin{remark} \label{remark1}
The Hastings-McLeod solution of (\ref{Painleve2}) with $\alpha = -1/2$
was introduced in \cite{HM}. It is chararacterized by the asymptotic
behavior $q(s) \sim \Ai(s)$ as $s \to +\infty$ where $\Ai$ is the usual
Airy function. For general $\alpha > -1$, $\alpha \neq -1/2$, the Hastings-McLeod solution of (\ref{Painleve2})
is characterized by the
asymptotic behavior $q(s) \sim (\alpha+\frac{1}{2})/s$ as $s \to +\infty$
and $q(s) \sim \sqrt{-s/2}$ as $s \to -\infty$,
see \cite[Chapter 11.7]{FIKN}. It was shown in \cite{CKV} that the
Hastings-McLeod solution has no poles on the real line and
so the system (\ref{diffsystem}) is well-defined for every $s \in \mathbb R$.

Note that $p_1 = q^2 + r + \frac{s}{2}$ satisfies
\[ p'' = 2 p^2 - sp + \frac{(p')^2 - \alpha^2}{2p} \]
and $p_2 = q^2 - r + \frac{s}{2}$ satisfies
\[ p'' = 2p^2 - sp + \frac{(p')^2 - (\alpha + 1)^2}{2p} \]
which are versions of the Painlev\'e XXXIV equation, see
\cite{FIKN}.

The functions $f_{\alpha}$ and $g_{\alpha}$ also satisfy
the following linear differential equation with respect to $s$
\begin{equation} \label{diffsystem2}
    \frac{\partial }{\partial s}
    \begin{pmatrix} f_{\alpha}(x;s) \\[5pt] g_{\alpha}(x;s) \end{pmatrix}
    = \begin{pmatrix} q & 1 \\[5pt]
    -x  & -q \end{pmatrix}
    \begin{pmatrix} f_{\alpha}(x;s) \\[5pt] g_{\alpha}(x;s) \end{pmatrix}
    \end{equation}
which together with (\ref{diffsystem}) constitutes the Lax pair for
Painlev\'e XXXIV, see \cite[page 175]{FIKN}. We will not use (\ref{diffsystem2})
in what follows.
In \cite{IKO}, a limiting kernel is obtained which is built out of
solutions to the Lax pair for Painlev\'e XXXIV as well. However,
the relevant solution to Painlev\'e XXXIV in this paper is
different from ours, and not related to the Hastings-McLeod
solution to Painlev\'e II.
\end{remark}

\begin{remark} \label{remark2}
The system (\ref{diffsystem}) can also be considered for complex
values of $x$. The functions $f_\alpha$ and $g_\alpha$ have
asymptotics in the complex $x$-plane given by
\begin{align} \label{falphaasymptotics}
    f_{\alpha}(x;s) & = \frac{1}{2 x^{1/4}} e^{\frac{1}{2}\pi i (\alpha+\frac{1}{2})}
    e^{-i(\frac{4}{3}x^{3/2}+sx^{1/2})} (1 + \bigO(x^{-1/2})) \\
    \label{galphaasymptotics}
    g_{\alpha}(x;s) & = \frac{x^{1/4}}{2 i} e^{\frac{1}{2}\pi i (\alpha+\frac{1}{2})}
    e^{-i(\frac{4}{3}x^{3/2}+sx^{1/2})} (1 + \bigO(x^{-1/2}))
\end{align}
uniformly as $x \to \infty$ in the sector $\varepsilon < \arg x <
2\pi - \varepsilon$ for any $\varepsilon > 0$. The positive real
line $\arg x = 0$ is an anti-Stokes line for the system
(\ref{diffsystem}), where the asymptotic behaviors
(\ref{falphaasymptotics}) and (\ref{galphaasymptotics}) are not
valid. As usual one has a two-term asymptotic approximation along
the anti-Stokes line which leads to
(\ref{realfalphaasymptotics}) and (\ref{realgalphaasymptotics}).
\end{remark}

\begin{remark}
Besides the eigenvalue correlation kernel, one might be interested
in the limiting distribution of the (re-scaled) smallest eigenvalue
of the ensemble (\ref{ensemble1}).  These distributions are given
by Fredholm determinants of the limiting kernels (\ref{integrablekernel}).
For general $\alpha$ we do not know if these kernels allow  Painlev\'e
expressions but for $\alpha = 0$ we can relate them to the
well-known Tracy-Widom distribution \cite{TW1}
\begin{equation} \label{TWdist}
    F_{\rm TW}(x) = \exp\left(- \int_x^{\infty} (y-x)^2 q(y) \, dy\right)
\end{equation}
where $q(y)$ is the Hastings-McLeod solution of Painlev\'e II.

Indeed, since $V$ is real analytic on $[0,\infty)$ we can
also consider $V$ on  $(-\delta, \infty)$ for some $\delta > 0$. For
small enough $\delta$ the equilibrium measure associated
with $V$ on $(-\delta, \infty)$ will not depend on $\delta$
and it will vanish like a square root at $0$.
Then for the ensemble (\ref{ensemble1}) with $\alpha = 0$
and defined on Hermitian matrices with
eigenvalues $\geq - \delta$ we have that the distribution
of the smallest eigenvalue tends to the Tracy-Widom distribution
(\ref{TWdist}) as $n \to \infty$. Here we assume for convenience that $N = n$;
otherwise under the assumption (\ref{eq-limits})
we find a shifted version of (\ref{TWdist}).
Moreover, denoting the smallest eigenvalue by $\lambda_1$ we have for $x>0$,
\begin{equation}
    \mathbb P(\lambda_1 > x) =
    \tilde{\mathbb P}(\lambda_1 > x \mid \lambda_1 > 0)
    = \frac{\tilde{\mathbb P}(\lambda_1 > x)}{\tilde{\mathbb P}(\lambda_1 > 0)}.
\end{equation}
Here $\mathbb P$ denotes the probability in the ensemble
(\ref{ensemble1}) on positive-definite matrices,
while $\tilde{\mathbb P}$ is the probability in the ensemble (\ref{ensemble1})
on Hermitian matrices with eigenvalues $\geq - \delta$.
Hence for an appropriate constant $c > 0$ we have if $\alpha = 0$
and $N = n$,
\begin{equation} \label{cutoffTW}
    \lim_{n \to \infty} \mathbb P(c n^{2/3} \lambda_1 > x)
    = \frac{F_{\rm TW}(-x)}{F_{\rm TW}(0)}.
\end{equation}
For $\alpha\neq 0$, the right-hand side of (\ref{cutoffTW})
will be different, but the scaling with $n^{2/3}$ in
the left-hand side will be the same.
\end{remark}

\section{A quadratic transformation}
In this section we relate the correlation kernels (\ref{kernel1}) for the random
matrix ensemble (\ref{ensemble1}) to the eigenvalue correlation kernel for
the random matrix ensemble (\ref{ensemble2}) with
\begin{equation} \label{defW}
    W(x) = \frac{1}{2} V(x^2), \qquad x \in \mathbb R
    \end{equation}
and parameter $\alpha \pm 1/2$ instead of $\alpha$. We use
$K_{n,N}^{(2\alpha \pm 1,W)}(x,y)$ to denote the correlation kernel.
Thus
\begin{equation}\label{kernel2}
    K_{n,N}^{(2\alpha\pm 1,W)}(x,y)= |x|^{\alpha \pm 1/2} |y|^{\alpha \pm 1/2}
    e^{-\frac{N}{2}W(x)}e^{-\frac{N}{2}W(y)}
    \sum_{j=0}^{n-1}P_j^{\supN}(x)P_j^{\supN}(y),
\end{equation}
where $P_j^{\supN}$ is the $j$-th degree orthonormal polynomial
with respect to the weight $|x|^{2\alpha \pm 1}e^{-NW(x)}$ on $\mathbb R$.

To distinguish this kernel from the kernel (\ref{kernel1}) that is
related to $V$ and parameter $\alpha$ we write in this section
\[ K_{n,N}(x,y) = K_{n,N}^{(\alpha, V)}(x,y). \]
The following result is not new, see e.g.\ \cite[Appendix B]{ADDV}
and  \cite[Exercises 4.4]{Forrester}. For the convenience of the
reader we have included a proof.

\begin{proposition} \label{kernel-relation}
Let $V$ be defined on $[0,\infty)$ and let $W$ be defined by {\rm (\ref{defW})}.
Assume that {\rm (\ref{asympVW})} is satisfied. Then for  $\alpha>-1$, we have
\begin{equation} \label{kernel-relation1}
    K_{n,N}^{(\alpha,V)}(x,y) =
    \frac{1}{2}(xy)^{-\frac{1}{4}}
    \left(K_{2n,2N}^{(2\alpha+1,W)}(\sqrt{x}, \sqrt{y}) + K_{2n,2N}^{(2\alpha+1,W)}(\sqrt{x}, -\sqrt{y})\right),
\end{equation}
while for $\alpha>0$, we have in addition
\begin{equation} \label{kernel-relation2}
    K_{n,N}^{(\alpha,V)}(x,y) = \frac{1}{2}(xy)^{-\frac{1}{4}}
    \left(K_{2n,2N}^{(2\alpha-1,W)}(\sqrt{x}, \sqrt{y})-K_{2n,2N}^{(2\alpha-1,W)}(\sqrt{x}, -\sqrt{y})\right).
\end{equation}
\end{proposition}

\begin{proof}
In the proof we use the abbreviations
\[ K_{n,N} = K_{n,N}^{(\alpha, V)}, \qquad
    K_{n,N}^+ = K_{n,N}^{(2\alpha+1,W)}, \qquad
    K_{n,N}^- = K_{n,N}^{(2\alpha-1,W)}. \]
We also write $p_n^{\supN}$ for the orthonormal polynomial of degree $n$ with
respect to the weight $w(x)=x^\alpha e^{-NV(x)}$ on $[0,\infty)$,
and $P_n^{\supN}$ and $Q_n^{\supN}$ for the orthonormal
polynomials with respect to the weights $w_1(x)=|x|^{2\alpha +
1}e^{-NW(x)}$ and $w_2(x)=|x|^{2\alpha -
1}e^{-NW(x)}$, respectively. Note that
$p_n^{\supN}$ and $P_n^{\supN}$ are well-defined for $\alpha>-1$,
while $Q_n^{\supN}$ is only defined for $\alpha>0$.

Since $w_1$ is an even weight on $\mathbb R$, the even degree
orthogonal polynomials are even and the odd degree orthogonal
polynomials are odd. Therefore there exists a polynomial $\widetilde{p}_n^{\supN}$
of degree $n$ such that
\begin{equation}\label{deftildepn}
P_{2n}^{(2N)}(x)=\widetilde p_n^{\supN}(x^2)
\end{equation}
For $k < n$, we then have
\begin{eqnarray}
    0&=&\int_{-\infty}^{\infty}P_{2n}^{(2N)}(x)x^{2k}|x|^{2\alpha +1}e^{-2NW(x)}dx\nonumber
    \\&=&2\int_0^{\infty}\widetilde p_n^{\supN}(x^2)x^{2k}x^{2\alpha +1}e^{-N V(x^2)}dx\nonumber\\
    &=&\int_0^{\infty}\widetilde p_n^{\supN}(s)s^{k}s^{\alpha}e^{-NV(s)}ds,
\end{eqnarray}
where we put $s = x^2$.
A similar calculation shows that
\begin{eqnarray}
    1 &= & \int_{-\infty}^{\infty} \left(P_{2n}^{(2N)}(x) \right)^2 |x|^{2\alpha +1}e^{-2NW(x)}dx\nonumber \\
    &=&\int_0^{\infty} \left(\widetilde p_n^{\supN}(s) \right)^2 s^{\alpha}e^{-NV(s)}ds,
\end{eqnarray}
Therefore $\widetilde p_n^{\supN}$ has the orthogonality conditions that characterize
the orthonormal polynomial $p_n^{\supN}$ and it follows that
\begin{equation} \label{P2n}
     P_{2n}^{(2N)}(x) = p_n^{(N)}(x^2).
\end{equation}
If $\alpha > 0$, we find in a similar way that
\begin{equation} \label{Q2n+1}
    Q_{2n+1}^{(2N)}(x) = x p_n^{(N)}(x^2).
\end{equation}

\bigskip

From (\ref{kernel2}), we now get for $x,y\geq 0$,
\begin{multline}\label{5-KnN1}
K_{2n,2N}^+(\sqrt x, \sqrt y)+ K_{2n,2N}^+(\sqrt x, -\sqrt
y)=(xy)^{\frac{\alpha}{2} + \frac{1}{4}}
e^{-\frac{N}{2} (V(x)+V(y))}\\
\times\ \left(\sum_{j=0}^{2n-1}P_j^{(2N)}(\sqrt x)P_j^{(2N)}(\sqrt
y)+
    \sum_{j=0}^{2n-1}P_j^{(2N)}(\sqrt x)P_j^{(2N)}(-\sqrt y)\right).
\end{multline}
Since $P_j^{(2N)}$ is even for even $j$ and
odd for odd $j$, we see that the odd $j$-terms in the two sums of (\ref{5-KnN1}) cancel
out, while the even $j$-terms are equal. We then obtain by
(\ref{P2n}) and (\ref{kernel1}) that
\begin{align*}
    &K_{2n,2N}^+(\sqrt x, \sqrt y)+ K_{2n,2N}^+(\sqrt x, -\sqrt y)\\
    &\qquad =2(xy)^{\frac{\alpha}{2} + \frac{1}{4}}e^{-\frac{N}{2}(V(x)+ V(y))}
    \sum_{j=0}^{n-1}P_{2j}^{(2N)}(\sqrt x)P_{2j}^{(2N)}(\sqrt y)\\
    &\qquad =2(xy)^{\frac{\alpha}{2} + \frac{1}{4}}e^{-\frac{N}{2}(V(x)+ V(y))}
    \sum_{j=0}^{n-1}p_{j}^{(N)}(x)p_{j}^{(N)}(y)\\
    &\qquad=2(xy)^{\frac{1}{4}}K_{n,N}(x,y).
\end{align*}
This proves (\ref{kernel-relation1}).

If $\alpha > 0$ and if we use (\ref{Q2n+1}) we obtain (\ref{kernel-relation2})
in a similar way.
\end{proof}

We will also connect the density $\psi_V$ of the equilibrium
measure for $V$ with the density $\psi_W$ of the equilibrium
measure for $W$. The equilibrium density $\psi_V$ minimizes
the energy (\ref{energy1}). It is characterized as the unique
probability density function on $[0, \infty)$ with the
property that for some constant $\ell_V$,
\begin{align}
2\int \log |x-y| \, \psi_V(y) \, dy - V(x) + \ell_V & =0 \qquad \mbox{ for }
x\in \supp \psi_V,  \label{varcond1a}\\
2\int \log |x-y| \, \psi_V(y) \, dy - V(x) + \ell_V & \leq 0 \qquad \mbox{ for }
x\in [0,\infty).  \label{varcond1b}
\end{align}

The density $\psi_W$ is the limiting mean eigenvalue density
for the matrix ensemble (\ref{ensemble2}) as $n,N\to\infty$ in such a way that
$n/N\to 1$. Then $\psi_W$ minimizes the weighted energy
\begin{equation} \label{energy2}
I_W(\psi)=\iint \log \frac{1}{|x-y|} \, \psi(x) \psi(y) \, dx dy
    + \int W(x) \, \psi(x) \, dx
\end{equation}
among all probability density functions $\psi$ on $\mathbb R$. The density
$\psi_W$
is uniquely characterized by the following variational conditions
for some constant $\ell_W$:
\begin{align}
2\int \log |x-y| \, \psi_W(y) \, dy - W(x) + \ell_W & =0 \qquad \mbox{ for }
x\in \supp \psi_W,  \label{varcond2a} \\
2\int \log |x-y| \, \psi_W(y) \, dy - W(x) + \ell_W & \leq 0 \qquad \mbox{ for }
x\in\mathbb R.  \label{varcond2b}
\end{align}

\begin{lemma} \label{lemma-eqdensity}
We have that
\[
    \psi_W(x) = |x|\psi_V(x^2) \qquad \mbox{for } x\in\mathbb R.
    \]
\end{lemma}
\begin{proof}
We first show that $|x|\psi_V(x^2)$ is indeed
a probability density. This follows from the positivity of $\psi_V$
and the fact that
\[
\int_{-\infty}^{\infty}|x|\psi_V(x^2)dx  = 2\int_0^{\infty}x\psi_V(x^2)dx
=\int_0^{\infty}\psi_V(s)ds=1.
\]

To prove part (i) of the lemma, it is now sufficient to show that $|x|\psi_V(x^2)$ satisfies
the variational conditions (\ref{varcond2a}) and (\ref{varcond2b}). We have
\begin{align*}
&\int_{-\infty}^{\infty}\log|x-y| \, |y|\psi_V(y^2) \, dy\\
&\qquad\qquad =
\int_{0}^{\infty}\log|x-y|\, y\psi_V(y^2) \, dy+
\int_{0}^{\infty}\log|x+y|\, y\psi_V(y^2)\, dy\\
&\qquad\qquad = \int_0^{\infty}\log|x^2-y^2|\, y\psi_V(y^2) \, dy\\
&\qquad\qquad =\frac{1}{2}\int_0^{\infty}\log|x^2-s| \, \psi_V(s) \, ds.
\end{align*}
This relation implies (\ref{varcond2a}) and
(\ref{varcond2b}) by virtue of (\ref{varcond1a}) and (\ref{varcond1b}).
Since these variational conditions uniquely characterize $\psi_W$
the lemma follows.
\end{proof}

In exactly the same way we find the following relation between
the equilibrium densities of the sets $S_V$ and $S_W$. Thus
$\omega_V$ is the unique probability density on $S_V$ that minimizes
the (unweighted) energy
\[ \iint \log \frac{1}{|x-y|} \, \omega(x) \omega(y) \, dx dy. \]

\begin{lemma} \label{lemma-omega}
Let $\omega_V$ be the equilibrium density of the set $S_V$ and let $\omega_W$
be the equilibrium density of $S_W = \{x \mid x^2\in S_V\}$. Then
\begin{equation}
\omega_W(x)=|x| \omega_V(x^2) \qquad \mbox{for } x\in S_W.
\end{equation}
\end{lemma}

\section{Proof of Theorem \ref{theorem11}}

The proof of Theorem \ref{theorem11} is based on the following
result for the matrix ensemble (\ref{ensemble2}) for the
case that $\psi_W$ vanishes quadratically at $0$.
There is a one-parameter family of limiting kernels that are connected with
the Hastings-McLeod solution of the Painlev\'e II equation $q'' = sq + 2q^3 - \alpha$.
We denote these kernels by $\mathbb K^{\rm crit,II}_{\alpha}(u,v;s)$
and we refer to (\ref{Kcritalpha1}) below for the precise form of $\mathbb K_{\alpha}^{\rm crit, II}(u,v;s)$.
The superscript II refers to the classification of singular points for the
matrix ensemble (\ref{ensemble2})  see \cite{DKMVZ1,KM}
according to which there are three cases
\begin{description}
\item[Singular case I]: Equality holds in (\ref{varcond2b}) for some
$x \in \mathbb R \setminus \supp \psi_W$. Any such point $x$ is then
a singular exterior point in the external field $W$.
\item[Singular case II]: $\psi_W$ vanishes at an interior point of its support.
Such a point is a singular interior point in the external field $W$.
\item[Singular case III]: $\psi_W$ vanishes at an edge point to higher order
than a square root. Such a point is a singular edge point in the external field $W$.
\end{description}

The following result was established in \cite{CKV}.

\begin{theorem} {\rm \textbf{(\cite[Theorem 1.2]{CKV})}} \label{theoremCKV}
Let $W$ be real analytic on $\mathbb R$ such that {\rm (\ref{asympVW})} holds.
Suppose that $\psi_W(0) = \psi_W'(0) = 0$ and $\psi_W''(0) > 0$, and assume that there
no other singular points besides $0$. Define constants
\begin{equation} \label{defc1Wc2W}
    c_{1,W} = \frac{\pi}{8} \psi_W''(0) > 0
    \qquad \mbox{ and } \qquad
    c_{2,W} = \frac{\pi}{c_{1,W}^{1/3}} \omega_{W}(0) > 0
\end{equation}
where $\omega_{W}$ is the density of the equilibrium measure of the support of $\psi_W$.
Let $n, N \to \infty$ such that
$ \lim\limits_{n,N \to \infty} n^{2/3} (n/N-1) = L \in \mathbb R$
exists and put $c = c_{1,W}$ and $s = c_{2,W} L$.
Then
\begin{equation} \label{scalinglimitW}
    \lim_{n,N \to \infty}
    \frac{1}{(cn)^{1/3}} K_{n,N}^{(\alpha,W)} \left( \frac{x}{(cn)^{1/3}}, \frac{y}{(cn)^{1/3}}\right)
        = \mathbb K_{\alpha}^{\rm crit,II}(x,y;s)
\end{equation}
uniformly for $x$, $y$ in compact subsets of $\mathbb R \setminus \{0\}$.
\end{theorem}

Now we are ready for the proof of Theorem \ref{theorem11} with constants
\begin{equation} \label{defc1Vc2V}
    c_{1,V} = \frac{\pi}{2} \lim_{x \to 0+} \left[x^{-1/2} \psi_V(x) \right]
    \qquad  \mbox{ and } \qquad
    c_{2,V} = \frac{2\pi}{c_{1,V}^{1/3}} \lim_{x \to 0+ } \left[\sqrt x \ \omega_{V}(x) \right],
\end{equation}
where $\omega_{V}$ is the density of the equilibrium measure of the support $S_V$
of $\psi_V$.

\begin{proof}[Proof of Theorem \ref{theorem11}.]
We define
\[ W(x) = \frac{1}{2} V(x^2), \qquad x \in \mathbb R. \]
Then by Lemma \ref{lemma-eqdensity} we have that $\psi_W(x) = |x| \psi_V(x^2)$,
which implies by the assumptions on $\psi_V$ near $0$, that
\begin{equation}
\psi_W(0)=\psi_W'(0)=0, \qquad
\psi_W''(0) = 2 \lim_{x \to 0+} \left[x^{-1/2} \psi_V(x)\right] > 0.
\end{equation}
It follows from Lemma \ref{lemma-eqdensity} that there are no other
singular points besides $0$ in the external field $W$ (in fact, the
absence of type I singular points follows from the proof of the
lemma). The conditions of Theorem \ref{theoremCKV} are therefore
satisfied. From (\ref{defc1Vc2V}), (\ref{defc1Wc2W}) and
Lemmas \ref{lemma-eqdensity} and \ref{lemma-omega} it follows that
\[ c_{1,V} = 2 c_{1,W}, \qquad
    \mbox{ and } \qquad c_{2,V} = 2^{2/3} c_{2,W}. \]
Then if $n, N \to \infty$ such that (\ref{eq-limits})
 and if $c = c_{1,V}$ and $s = c_{2,V}L$,
 we get from from Theorem \ref{theoremCKV} that
\begin{equation} \label{scalinglimitW3}
    \lim_{n,N \to \infty} \frac{1}{(cn)^{1/3}}
    K_{2n,2N}^{(2\alpha+1,W)} \left( \frac{x}{(cn)^{1/3}}, \frac{y}{(cn)^{1/3}}\right) =
    \mathbb K_{\alpha+ \frac{1}{2}}^{\rm crit,II}(x,y;s),
    \end{equation}
and if $\alpha > 0$ we also have
\begin{equation} \label{scalinglimitW4}
    \lim_{n,N \to \infty} \frac{1}{(cn)^{1/3}}
    K_{2n,2N}^{(2\alpha-1,W)} \left( \frac{x}{(cn)^{1/3}}, \frac{y}{(cn)^{1/3}}\right) =
    \mathbb K_{\alpha-\frac{1}{2}}^{\rm crit,II}(x,y;s),
    \end{equation}
Then straightforward calculations based on (\ref{kernel-relation1})
and (\ref{scalinglimitW3}) show that the limit (\ref{limitkernel}) exists
with
\begin{equation}
    \mathbb K_{\alpha}^{\rm soft/hard}(x,y;s) =
       \frac{1}{2}(xy)^{-1/4}\left(\mathbb K_{\alpha +\frac{1}{2}}^{\rm crit, II}(\sqrt x,\sqrt y;s)
       +\mathbb K_{\alpha +\frac{1}{2}}^{\rm crit, II}(\sqrt x,-\sqrt y;s)\right),
\label{scalinglimitV1} \end{equation}
while if $\alpha > 0$, then from (\ref{kernel-relation2}) and
(\ref{scalinglimitW4}) we also get
\begin{equation}
    \mathbb K_{\alpha}^{\rm soft/hard}(x,y;s) =
        \frac{1}{2}(xy)^{-1/4}\left(\mathbb K_{\alpha-\frac{1}{2}}^{\rm crit, II}(\sqrt x,\sqrt y;s)
        - \mathbb K_{\alpha-\frac{1}{2}}^{\rm crit, II}(\sqrt x,-\sqrt y;s)\right).
    \label{scalinglimitV2} \end{equation}
\end{proof}

\section{Proof of Theorem \ref{theorem12}}

To prove Theorem \ref{theorem12}, we recall the definition of the kernel
$\mathbb K_{\alpha}^{\rm crit,II}$ from \cite{CKV}. It uses the Hastings-McLeod
solution $q=q_{\alpha}$ of the Painlev\'e II equation
\begin{equation} \label{PainleveII}
    q'' = sq + 2q^3 - \alpha.
\end{equation}

Given $\alpha > -1/2$ and $s \in \mathbb R$, we abbreviate $q =
q_{\alpha}(s)$ and $r = q_{\alpha}'(s)$, and we let
$\begin{pmatrix}\Phi_{\alpha,1}(z;s) \\
\Phi_{\alpha,2}(z;s) \end{pmatrix}$ be the unique solution of
the equation
\begin{equation} \label{diffeqPhi}
\frac{d}{dz}\begin{pmatrix}
\Phi_{\alpha,1}\\
\Phi_{\alpha,2}
\end{pmatrix}=\begin{pmatrix}
            -4i z^2-i(s+2q^2) & 4z q+2ir+\alpha /z  \\
            4 z q -2i r+\alpha /z   & 4i z^2+i(s+2q^2)
        \end{pmatrix}\begin{pmatrix}
\Phi_{\alpha,1}\\
\Phi_{\alpha,2}
\end{pmatrix},
\end{equation}
with asymptotics
\begin{equation}\label{asymptoticsPhi}
    e^{i(\frac{4}{3} z^3+sz)}
    \begin{pmatrix}\Phi_{\alpha,1}(z;s)\\
    \Phi_{\alpha,2}(z;s)\end{pmatrix}
    =\begin{pmatrix}1\\0\end{pmatrix}+\bigO(z^{-1}),
\end{equation}
uniformly as $z\to\infty$ in the sector $ \varepsilon<\arg
z <\pi -\varepsilon$ for any $\varepsilon>0$. Since $q_\alpha$
has no real poles for $\alpha>-1/2$, $\Phi_{\alpha,1}$ and
$\Phi_{\alpha,2}$ are well-defined for real $s$. The functions
$\Phi_{\alpha,1}$ and $\Phi_{\alpha,2}$ extend to analytic
functions on $\mathbb C \setminus (-i\infty, 0]$, with branch
points in $0$. We denote these extensions also by
$\Phi_{\alpha,1}$ and $\Phi_{\alpha,2}$. In the kernel
$\mathbb K_{\alpha}^{\rm crit, II}$ we need their
values on the real line.
Indeed, we have for real $x$, $y$, and $s$,
\begin{equation} \label{Kcritalpha1}
        \mathbb K_\alpha^{\rm crit, II}(x,y;s)=  -e^{\frac{1}{2}\pi
        i\alpha[\sgn(x)+\sgn(y)]}
        \frac{\Phi_{\alpha,1}(x;s) \Phi_{\alpha,2}(y;s) -
        \Phi_{\alpha,1}(y;s)\Phi_{\alpha,2}(x;s)}{2\pi i(x-y)}.
\end{equation}

We first rewrite $\mathbb K_{\alpha}^{\rm crit, II}$ in
a `real' form, using the functions (see also \cite{BI2}),
\begin{equation} \label{defFalpha}
    \begin{pmatrix} F_{\alpha,1}(x;s) \\ F_{\alpha,2}(x;s) \end{pmatrix}
    =  e^{\frac{1}{2} \pi i \alpha \sgn(x)} \frac{1}{2} \begin{pmatrix} 1 & 1 \\ -i & i \end{pmatrix}
        \begin{pmatrix} \Phi_{\alpha,1}(x;s) \\ \Phi_{\alpha,2}(x;s) \end{pmatrix},
            \qquad x \in \mathbb R \setminus \{0\}.
            \end{equation}
Then we get from (\ref{Kcritalpha1}) that
\begin{equation} \label{Kcritalpha2}
        \mathbb K_\alpha^{\rm crit, II}(x,y;s) =
          \frac{F_{\alpha,1}(x;s) F_{\alpha,2}(y;s) -
        F_{\alpha,1}(y;s) F_{\alpha,2}(x;s)}{\pi (x-y)}
\end{equation}
and from (\ref{diffeqPhi}) we get the differential equation
\begin{equation} \label{diffeqF}
\frac{d}{dx} \begin{pmatrix}
F_{\alpha,1}\\
F_{\alpha,2}
\end{pmatrix}=\begin{pmatrix}
            4x q +\alpha /x   & 4x^2+ s+2q^2 + 2r \\
            -4x^2- s -2q^2 + 2r & -4x q -\alpha /x
        \end{pmatrix}\begin{pmatrix}
F_{\alpha,1}\\
F_{\alpha,2}
\end{pmatrix}.
\end{equation}

The following properties are crucial.
\begin{lemma} \label{lemma-even-odd}
Both $F_{\alpha,1}(x;s)$ and $F_{\alpha,2}(x;s)$ are real for real $x$ and $s$.
In addition, $F_{\alpha,1}$ is an even function and $F_{\alpha,2}$ is an odd function,
and they have the following asymptotic behaviors
\begin{align} \label{F1asymptotics}
F_{\alpha,1}(x;s) & =  \cos\left(\frac{4}{3}x^{3}+sx -
\frac{1}{2}\pi \alpha\right)
     + \bigO(x^{-1}), \\
    \label{F2asymptotics}
    F_{\alpha,2}(x;s) & = -
        \sin\left(\frac{4}{3}x^{3}+sx - \frac{1}{2}\pi \alpha\right)
    + \bigO(x^{-1}),
\end{align}
as $x \to +\infty$.
\end{lemma}
\begin{proof}
The proof is based on the symmetries of the Riemann-Hilbert problem
that characterizes $\begin{pmatrix} \Phi_{\alpha,1} \\ \Phi_{\alpha,2} \end{pmatrix}$.

Let $\Gamma = \bigcup_j \Gamma_j$ be the contour consisting
of four straight rays oriented to infinity, where
\[
    \Gamma_1:\arg z=\frac{\pi}{6}, \qquad \Gamma_2:\arg z=\frac{5\pi}{6}, \qquad
    \Gamma_3:\arg z=-\frac{5\pi}{6}, \qquad \Gamma_4:\arg z=-\frac{\pi}{6}
\]
see Figure \ref{contourGamma}. The contour $\Gamma$ divides the complex plane
into four sectors $S_j$, $j=1,2,3,4$ as shown in Figure \ref{contourGamma}.

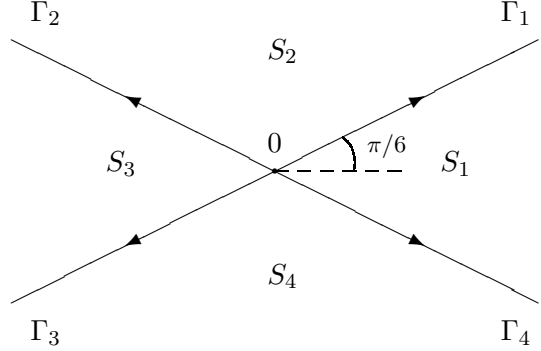
\begin{figure}[t]
\begin{center}
    \setlength{\unitlength}{1truemm}
    \begin{picture}(100,48.5)(0,2.5)
        \put(72,27.5){\small $S_1$}
        \put(49,42.5){\small $S_2$}
        \put(27.5,27.5){\small $S_3$}
        \put(49,12.5){\small $S_4$}

        \put(80,47.5){\small $\Gamma_1$}
        \put(17.5,47.5){\small $\Gamma_2$}
        \put(17.5,5){\small $\Gamma_3$}
        \put(80,5){\small $\Gamma_4$}

        \put(50,27.5){\thicklines\circle*{.8}}
        \put(49,30){\small 0}
        \multiput(50,27.5)(3,0){6}{\line(1,0){1.75}}
        \qbezier(60.5,27.5)(61,30.5)(59,32)
        \put(62.25,30){\scriptsize $\pi/6$}

        \put(15,10){\line(2,1){70}}
        \put(15,45){\line(2,-1){70}}
        \put(70,37.5){\thicklines\vector(2,1){.0001}}
        \put(70,17.5){\thicklines\vector(2,-1){.0001}}
        \put(30,37.5){\thicklines\vector(-2,1){.0001}}
        \put(30,17.5){\thicklines\vector(-2,-1){.0001}}
    \end{picture}
    \caption{The contour $\Gamma$ consisting of four straight rays oriented to infinity.}
    \label{contourGamma}
\end{center}
\end{figure}

For $\alpha > -1/2$ and $s \in \mathbb R$ we look for
$\Psi_{\alpha}(z;s)$
that satisfies the following Riemann-Hilbert problem.
\begin{itemize}
\item[(a)] $\Psi_{\alpha} : \mathbb C \setminus \Gamma \to \mathbb C^{2\times 2}$
    is analytic in $\mathbb C\setminus\Gamma$.
\item[(b)] $\Psi_{\alpha}$ satisfies the following jump
    relations on $\Gamma \setminus \{0\}$,
    \begin{align}
        \label{RHP Psi: b1}
        \Psi_{\alpha,+}(z;s) &= \Psi_{\alpha,-}(z;s)
            \begin{pmatrix}
                1 & 0 \\
                e^{-\pi i \alpha} & 1
            \end{pmatrix},
            \qquad \mbox{for $z\in\Gamma_1$,}
        \\[1ex]
        \label{RHP Psi: b2}
        \Psi_{\alpha,+}(z;s) &= \Psi_{\alpha,-}(z;s)
            \begin{pmatrix}
                1 & 0 \\
                -e^{\pi i \alpha} & 1
            \end{pmatrix},
        \qquad \mbox{for $z\in\Gamma_2$,}
        \\[1ex]
        \label{RHP Psi: b3}
            \Psi_{\alpha,+}(z;s) &= \Psi_{\alpha,-}(z;s)
            \begin{pmatrix}
                1 & e^{-\pi i \alpha} \\
                0 & 1
            \end{pmatrix},
            \qquad \mbox{for $z\in\Gamma_3$,}
        \\[1ex]
        \label{RHP Psi: b4}
            \Psi_{\alpha,+}(z;s) &= \Psi_{\alpha,-}(z;s)
            \begin{pmatrix}
                1 & -e^{\pi i \alpha} \\
                0 & 1
            \end{pmatrix},
            \qquad \mbox{for $z\in\Gamma_4$.}
    \end{align}
\item[(c)] $\Psi_{\alpha}$ has the following behavior at infinity,
    \begin{equation}\label{RHP Psi: c}
        \Psi_{\alpha}(z;s)=(I+\bigO(1/z))e^{-i(\frac{4}{3}z^3+sz)\sigma_3},
        \qquad \mbox{as $z\to\infty$.}
    \end{equation}
    Here $\sigma_3 = \left(\begin{smallmatrix} 1 & 0 \\ 0 & -1 \end{smallmatrix}\right)$
    denotes the third Pauli matrix.
\item[(d)] $\Psi_{\alpha}$ has the following behavior near the origin. If $\alpha<0$,
    \begin{equation}\label{RHP Psi: d1}
        \Psi_{\alpha}(z;s)=
        \bigO\begin{pmatrix}
            |z|^\alpha & |z|^\alpha \\
            |z|^\alpha & |z|^\alpha
        \end{pmatrix},
        \qquad \mbox{as $z\to 0$,}
    \end{equation}
    and if $\alpha\geq 0$,
    \begin{equation}\label{RHP Psi: d2}
        \Psi_{\alpha}(z;s)=
        \begin{cases}
            \bigO\begin{pmatrix}
                |z|^{-\alpha} & |z|^{-\alpha} \\
                |z|^{-\alpha} & |z|^{-\alpha}
            \end{pmatrix},
                & \mbox{as $z\to 0,\, z \in S_1\cup S_3$,}
            \\[3ex]
            \bigO\begin{pmatrix}
                |z|^{\alpha} & |z|^{-\alpha} \\
                |z|^{\alpha} & |z|^{-\alpha}
            \end{pmatrix},
                & \mbox{as $z\to 0,\, z \in S_2$,}
            \\[3ex]
            \bigO\begin{pmatrix}
                |z|^{-\alpha} & |z|^{\alpha} \\
                |z|^{-\alpha} & |z|^{\alpha}
            \end{pmatrix},
                & \mbox{as $z\to 0,\, z \in S_4$.}
        \end{cases}
    \end{equation}
\end{itemize}
It was shown in \cite{CKV} that the Riemann-Hilbert problem for
$\Psi_{\alpha}$ has a unique solution and that
\begin{equation}\label{definition: Phi_alpha}
    \begin{pmatrix} \Phi_{\alpha,1}(z;s) \\ \Phi_{\alpha,2}(z;s) \end{pmatrix}
    =
    \begin{cases}
        \Psi_{\alpha}(z;s)
            \begin{pmatrix}
                1 \\
                e^{-\pi i \alpha}
            \end{pmatrix}, & \mbox{for $z\in S_1$} \\[3ex]
        \Psi_{\alpha}(z;s)
            \begin{pmatrix}
                1 \\
                0
            \end{pmatrix}, & \mbox{for $z\in S_2$,} \\[2ex]
        \Psi_{\alpha}(z;s)
            \begin{pmatrix}
                1  \\
                e^{\pi i \alpha}
            \end{pmatrix}, & \mbox{for $z\in S_3$,} \\[3ex]
        \Psi_{\alpha}(z;s)
            \begin{pmatrix}
                0 \\
                e^{-\pi i \alpha}
            \end{pmatrix}, & \mbox{for $z \in S_4$, $\Re z > 0$,} \\[3ex]
        \Psi_{\alpha}(z;s)
            \begin{pmatrix}
                0  \\
                e^{\pi i \alpha}
            \end{pmatrix}, & \mbox{for $z \in S_4$, $\Re z < 0$.}
    \end{cases}
\end{equation}

It is easy to check that $\sigma_1\overline{\Psi_{\alpha}(\overline{z};s)}
\sigma_1$ and $\sigma_1 \Psi_{\alpha}(-z;s) \sigma_1$ with
$\sigma_1= \left(\begin{smallmatrix}
    0 & 1\\
    1 & 0
\end{smallmatrix}\right)$ also satisfy the conditions of the
Riemann-Hilbert problem for $\Psi_{\alpha}$. Therefore by the
uniqueness of the solution we have
\begin{equation} \label{symmetries2}
    \Psi_{\alpha}(z;s) =
        \sigma_1\overline{\Psi_{\alpha}(\overline{z}; s)}\sigma_1 =
        \sigma_1 \Psi_{\alpha}(-z;s) \sigma_1.
        \end{equation}
From the first equality in (\ref{symmetries2}) it follows (as
noted in Remark 2 of \cite{CKV}) that
\begin{equation} \label{symmetries}
    e^{\frac{1}{2}\pi i \alpha \sgn(x)} \Phi_{\alpha,2}(x;s) =
    \overline{e^{\frac{1}{2} \pi i \alpha \sgn(x)} \Phi_{\alpha,1}(x;s)},
    \qquad \mbox{for } x \in \mathbb R\setminus\{0\} \mbox{ and } s \in \mathbb R,
\end{equation}
which shows that $F_{\alpha,1}(x;s)$ and $F_{\alpha,2}(x;s)$ are real,
since we obtain from (\ref{defFalpha}) and (\ref{symmetries}) that
\begin{equation} \label{symmetries3}
\begin{aligned}
    F_{\alpha,1}(x;s) & = \Re \left[e^{\frac{1}{2} \pi i \alpha \sgn(x)} \Phi_{\alpha,1}(x;s) \right], \\
    F_{\alpha,2}(x;s) & = \Im \left[e^{\frac{1}{2} \pi i \alpha \sgn(x)} \Phi_{\alpha,1}(x;s) \right].
\end{aligned}
\end{equation}
From (\ref{RHP Psi: c}), (\ref{definition: Phi_alpha}), and
(\ref{symmetries3}) we obtain the asymptotic behaviors
(\ref{F1asymptotics}) and (\ref{F2asymptotics}).

The symmetry $\Psi_{\alpha}(z;s) = \sigma_1 \Psi_{\alpha}(-z;s) \sigma_1$
yields by (\ref{definition: Phi_alpha}) that
\[ e^{\frac{1}{2} \pi i \alpha \sgn(x)} \Phi_{\alpha,2}(x;s) =
    e^{\frac{1}{2} \pi i \alpha \sgn(-x)} \Phi_{\alpha,1}(-x;s),
    \qquad \mbox{ for } x \in \mathbb R \setminus \{0\}, \]
which in view of (\ref{symmetries}) and (\ref{symmetries3}) shows that $F_{\alpha,1}$
is even and $F_{\alpha,2}$ is odd.
\end{proof}

Now we can give the proof of Theorem \ref{theorem12}.

\begin{proof}[Proof of Theorem \ref{theorem12}.]
From (\ref{scalinglimitV1}) and (\ref{Kcritalpha2}), we get
\begin{multline}
    \mathbb K_{\alpha}^{\rm soft/hard}(x,y;s)  \\
    =
    \frac{1}{2}(xy)^{-1/4}
    \left( \frac{F_{\alpha+\frac{1}{2},1}(\sqrt x) F_{\alpha+\frac{1}{2},2}(\sqrt y) -
        F_{\alpha+\frac{1}{2},1}(\sqrt y) F_{\alpha+\frac{1}{2},2}(\sqrt x)}{\pi(\sqrt x-\sqrt y)}
         \right. \\
         \left. + \frac{F_{\alpha+\frac{1}{2},1}(\sqrt x) F_{\alpha+\frac{1}{2},2}(-\sqrt y) -
        F_{\alpha+\frac{1}{2},1}(-\sqrt y) F_{\alpha+\frac{1}{2},2}(\sqrt x)}{\pi(\sqrt x+\sqrt y)}
        \right).
\end{multline}
Putting the fractions onto a common denominator, and using the even/odd properties
of Lemma \ref{lemma-even-odd}, we find
\begin{multline}
    \mathbb K_{\alpha}^{\rm soft/hard}(x,y;s)  \\
    = \frac{1}{\pi} (xy)^{-1/4}
        \left( \frac{F_{\alpha+\frac{1}{2},1}(\sqrt x) \sqrt y F_{\alpha+\frac{1}{2},2}(\sqrt y)
            - F_{\alpha+\frac{1}{2},1}(\sqrt y) \sqrt x F_{\alpha+\frac{1}{2},2}(\sqrt x)}
            {x-y} \right)
\end{multline}
which leads to (\ref{integrablekernel}) with
\begin{equation} \label{deffalpha}
     f_{\alpha}(x;s) =  x^{-1/4} F_{\alpha+\frac{1}{2},1}(\sqrt x;s),
     \end{equation}
and
\begin{equation} \label{defgalpha}
    g_{\alpha}(x;s) = x^{1/4} F_{\alpha+\frac{1}{2},2}(\sqrt x;s).
    \end{equation}
The system of differential equations (\ref{diffsystem}) follows from
(\ref{diffeqF}), (\ref{deffalpha}), and (\ref{defgalpha}) and the
asymptotic behavior
(\ref{realfalphaasymptotics})-(\ref{realgalphaasymptotics}) follows
from (\ref{F1asymptotics})-(\ref{F2asymptotics}) and
(\ref{deffalpha})-(\ref{defgalpha}).

This completes the proof of Theorem \ref{theorem12}.
\end{proof}

\end{document}